\documentclass[prl,aps,twocolumn,epsfig,superscriptaddress,showpacs]{revtex4}
\usepackage{bm}
\usepackage{amsfonts}
\usepackage[dvips]{graphicx}
\usepackage{mathrsfs}
\usepackage[intlimits]{amsmath}
\usepackage[colorlinks, citecolor=red]{hyperref}

\begin{document}

\title{Exploring Quantum Contextuality to Generate True Random Numbers}
\author{D.-L. Deng$^{1,2}$, C. Zu$^{1}$, X.-Y. Chang$^{1}$, P.-Y. Hou$^{1}$, H.-X. Yang$^{1}$, Y.-X. Wang$^{1}$,
L.-M. Duan}
\affiliation{Center for Quantum Information, IIIS, Tsinghua University, Beijing,
China}
\affiliation{Department of Physics, University of Michigan, Ann Arbor, Michigan
48109, USA}

\begin{abstract}
Random numbers represent an indispensable resource for many
applications. A recent remarkable result is the realization that non-locality in quantum
mechanics can be used to certify genuine randomness through Bell's theorem,
producing reliable random numbers in a device independent way. Here, we explore the contextuality aspect of quantum mechanics and show that true random numbers can be generated using only single qutrit (three-state systems) without entanglement and non-locality.
 In particular, we show that any observed violation of the Klyachko-Can-Binicioglu-Shumovsky
(KCBS) inequality [Phys.
Rev. Lett. \textbf{101}, 20403 (2008)] provides a positive lower bound on genuine
randomness. As a proof-of-concept experiment, we demonstrate with photonic
qutrits that at least $5246$ net true random numbers are generated with
a confidence level of $99.9\%$.
\end{abstract}


\maketitle

Random numbers are widely used in algorithms and technology~\cite%
{1982Yao,1981Knuth}. However, generation of genuine randomness is a
challenging task~\cite{2010Pironio}. Mathematically, randomness means
unpredictability~\cite{2011Abbott, Chaitin-Book}. Thus, in principle, random
numbers can never be generated by a classical device since any classical
system bears a deterministic description. Consequently, random numbers
generated by a classical device can always be attributed to a lack of
knowledge about the device. If we know all the information of the device, in
principle we can predict all the results of any operation on this device.
Unlike classic systems, quantum theory is intrinsically random. It is
natural to think about generating random numbers via a quantum device. In
fact, various quantum random number generators (QRNGs) have already been
reported. Significant examples include those based on the decay of
radioactive nucleus~\cite{1956Isida}, beam splitters~\cite%
{1990Svozil,1994Rarity, 2000Jennewein}, entangled photon pairs~\cite{2004Ma}
and amplified quantum vacuum~\cite{2011Jofre}. However, in real experiment
the intrinsic randomness of these QRNGs is inevitably mixed-up with an
apparent randomness due to noise or lack of control of the experiment. In
other words, the randomness generated by these QRNGs cannot be unequivocally
certified or quantified. This will jeopardize some applications of
randomness, especially cryptographic applications. A breakthrough was made
by Colbeck~\cite{2007Colbeck} and subsequently developed by Pironio \textit{%
et al}~\cite{2010Pironio,correction}. The basic idea is to use the non-local
correlation of quantum states to generate certified private randomness. More
specifically, Bell's theorem can be used to certify genuine randomness. In
Ref.~\cite{2010Pironio}, taking the Clauser-Horn-Shimony-Holt (CHSH)
inequality~\cite{1969Clauser} as an example, Pironio \textit{et al}
demonstrated for the first time this important idea with a proof-of-concept
experiment using entangled trapped ions. A more recent work in this direction is Ref.~\cite{2012VV}.

Here, we introduce a new method to generate true random numbers in
single-qutrit systems through exploration of the Kochen-Specker (KS)
theorem. Generation of randomness by this method does not rely on the
costly quantum resource of entanglement, which significantly simplifies its
experimental realization. The Kochen-Specker theorem~\cite%
{1960Specker,1966Bell,1967Kochen} states that no non-contextual hidden
variable model (NCHVM) can reproduce the prediction of quantum mechanics, or
simply put, quantum mechanics is contextual. In recent years, extensive
works on quantum contextuality have been done, including both theoretical
analyses~\cite{2008KCBS,1990Peres,1990Mermin,2008Cabello,2012Yu} and
experiment demonstrations~\cite%
{2000Michler,2009Kirchmair,2009Bartosik,2009Amselem,2010Moussa,2011Lapkiewicz,2012Zu}%
. All the experimental results favor quantum mechanics and hence rule out
the NCHVM. Here, we exploit this theorem from a new angle and show that it
can be used  to generate genuine randomness. To this end, we explore a KS
inequality introduced recently by Klyachko, Can, Binicioglu and Shumovsky
(KCBS)~\cite{2008KCBS}, and show that any observed violation of the KCBS
inequality leads to a positive lower bound on the randomness produced by the
quantum device. Furthermore, as a proof-of-concept experiment, we
demonstrate this new method with photonic qutrits by showing that at least $%
5246$ net true random numbers are generated with a confidence
level of $99.9\%$. %As the scheme does not require entangling operations and
%therefore can operate with a high speed, it allows us to experimentally
%generate \textit{net} true random numbers.

To be specific, we consider a single qutrit system and five two-outcome
measurements $A_{i}$ $(i=1,2,3,4,5)$. Denoting the outcome of the the
corresponding measurement $A_{i}$ as $a_{i}$ ($a_{i}=0,1$), the KCBS
inequality can be rewritten as~\cite{2008KCBS, 2011Lapkiewicz}:
\begin{equation}
L\equiv \sum_{(i,j)\in \mathcal{S}}[P(a_{i}\neq
a_{j}|A_{i}A_{j})-P(a_{i}=a_{j}|A_{i}A_{j})]\leq 3,  \label{KCBS-Ineq}
\end{equation}%
where $\mathcal{S}=\{(1,2),(2,3),(3,4),(4,5),(5,1)\}$ represents the set of
pairs of compatible (commutable) measurements, and $P(a_{i}\neq
a_{j}|A_{i}A_{j})$ ($P(a_{i}=a_{j}|A_{i}A_{j})$) is respectively the
probability that $a_{i}\neq a_{j}$ ($a_{i}=a_{j}$) when the measurement
setting $(A_{i},A_{j})$ is chosen. The inequality~(\ref{KCBS-Ineq}) is
satisfied by any NCHVM. In quantum mechanics, however, this inequality can
be violated for certain measurements performed on a specific state and the
maximal violation is $4\sqrt{5}-5\approx 3.944$ \cite{2008KCBS}. An
experimental violation has been reported recently in Ref.~\cite%
{2011Lapkiewicz}. For our purpose to relate the KCBS violation to
the generation of randomness, we run the experiment $k$ times in succession.
The measurement choice $(A_{i},A_{j})$ for each trial is generated by a
computer through an identical and independent probability distribution $%
P(A_{i}A_{j})$ ($(i,j)\in \mathcal{S}$). Denoting the input string as $%
\mathcal{I}=(A_{i_{1}},A_{j_{1}};\cdots ;A_{i_{k}},A_{j_{k}})$ and the
corresponding output string as $\mathcal{O}=(a_{i_{1}},a_{j_{1}};\cdots
;a_{i_{k}},a_{j_{k}})$, the estimated KCBS violation can be obtained from
the observed data as
\begin{equation}
\hat{L}=\frac{1}{k}\sum_{(i,j)\in \mathcal{S}}[N(a_{i}\neq
a_{j}|A_{i}A_{j})-N(a_{i}=a_{j}|A_{i}A_{j})]/P(A_{i}A_{j}),
\label{Estimator-KCBS}
\end{equation}%
where $N(a_{i}\neq a_{j}|A_{i}A_{j})$ ($N(a_{i}=a_{j}|A_{i}A_{j})$) denotes
respectively the number of trials with unequal (equal) measurement outcomes
under the measurement setting $(A_{i},A_{j})$.

\begin{figure}[tbp]
%Requires \usepackage{graphicx}
\includegraphics[width=7.5cm]{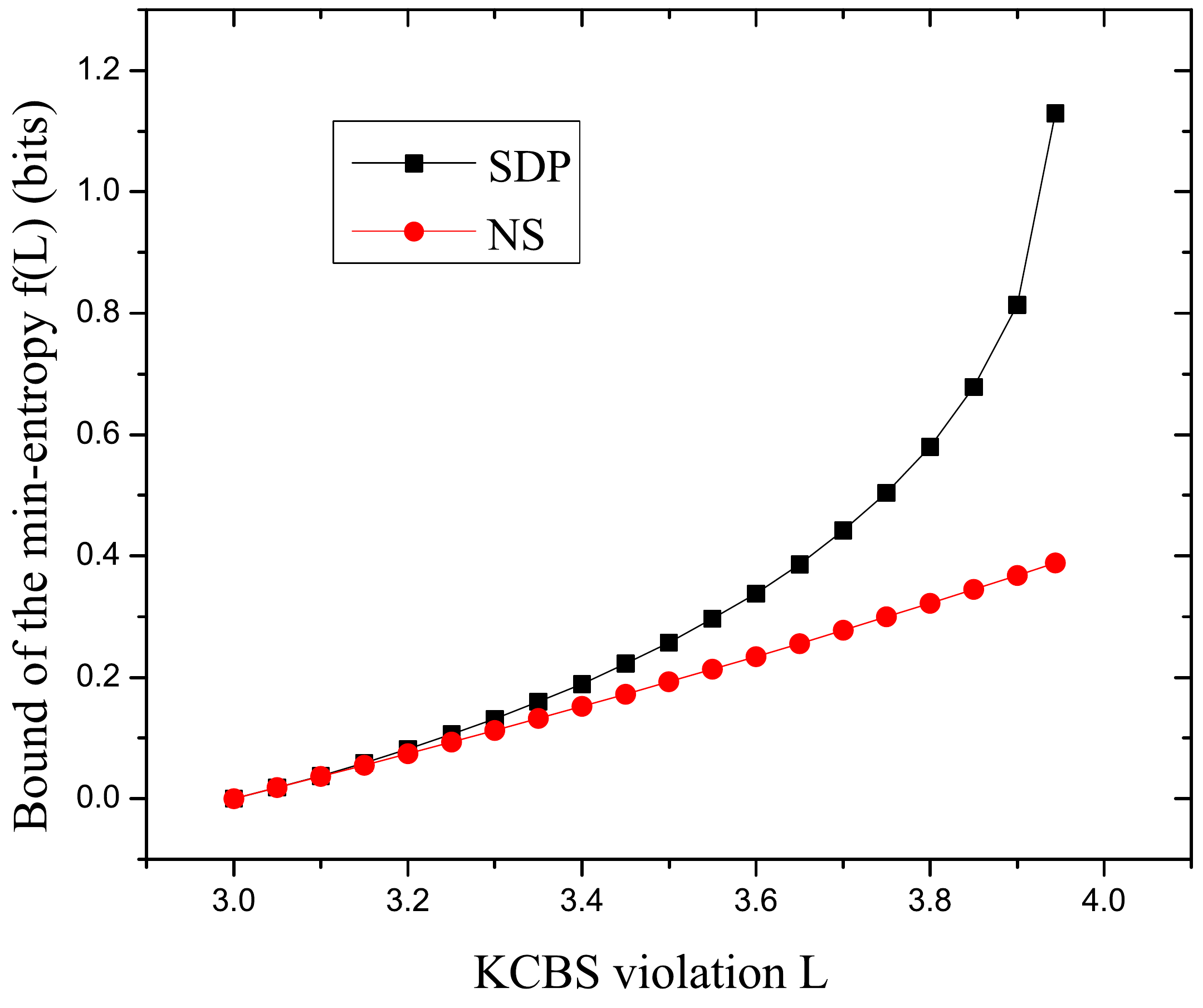}\newline
\caption{The min-entropy bound $f(L)$ versus different levels $L$ of the
KCBS violation. The black-square line is obtained through optimization based
on the semi-definite programming (SDP) assuming validity of quantum
mechanics. We can relax this assumption and do not assume the formalism of
quantum mechanics, but only require that for two compatible (commutable)
observables, a measurement on one observable does not change the marginal
probability distribution of measurement outcomes of the other observable.
This corresponds to the no signaling (NS) condition for bi-partite system
and we still call it the NS condition. The red-dotted line corresponds to an
analytical lower-bound $f(L)=-\log_2(1.75-L/4)$ obtained under the NS
condition only (see the supporting information for a detailed
derivation). }
\label{Minentropy}
\end{figure}

\begin{figure}[tbp]
%Requires \usepackage{graphicx}
\includegraphics[width=8.6cm]{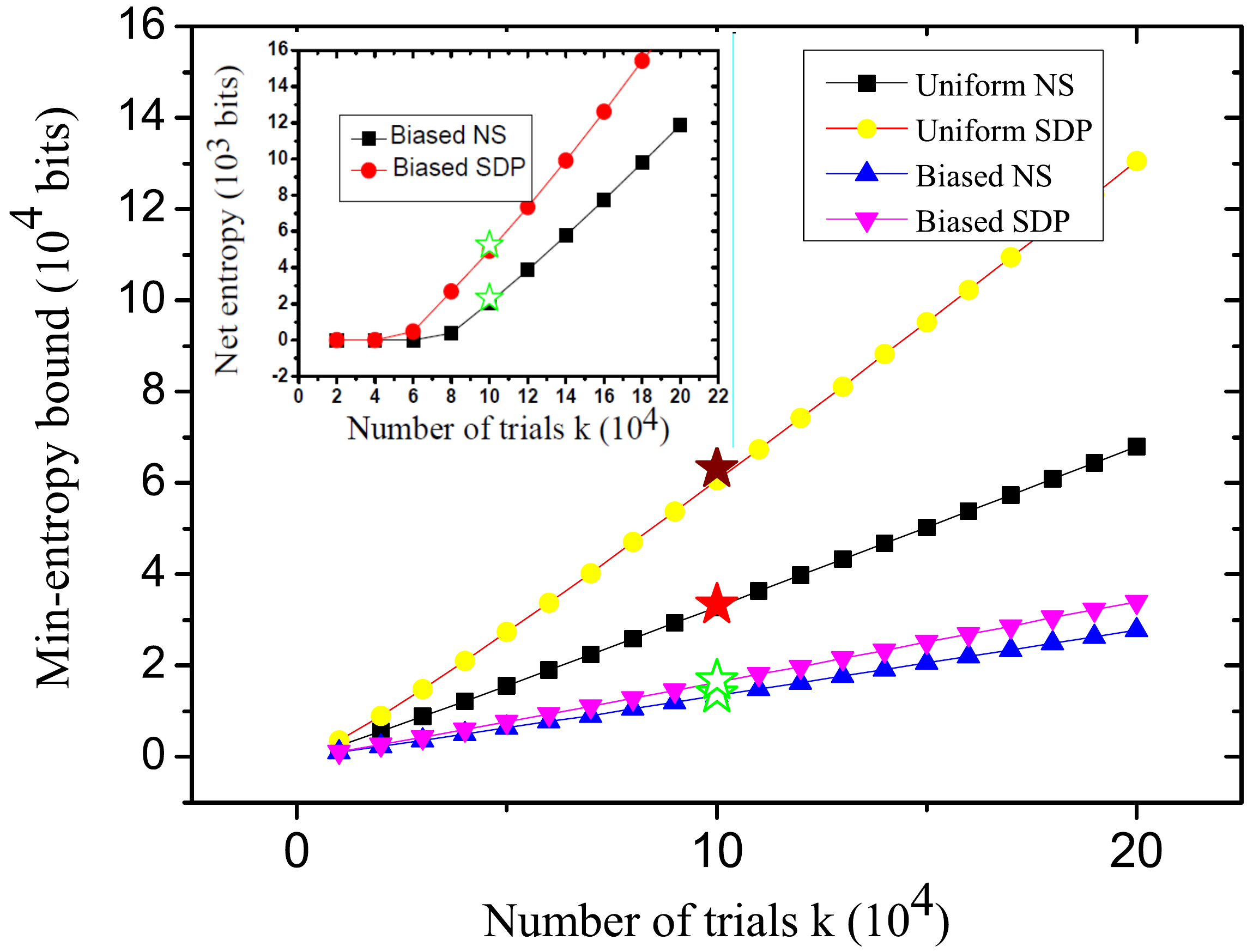}\newline
\caption{The min-entropy bound $kf(\mathcal{L}_m-\epsilon)-\log_2\frac{1}{\delta}$ versus the
number of trials $k$ assuming that the observed KCBS violation lies within the interval $3.9=\mathcal{L}_m\leq \hat{L}<\mathcal{L}_{m+1}=4\sqrt{5}-5\approx3.944$ with non-negligible probability $\delta$. Here the parameters are chosen as $\delta=0.001$ and $\epsilon'=0.01$.
 We take two different distributions for the input pairs $(A_i,A_j)$: a uniform
distribution with $P(A_iA_j)=1/5$ $(i,j)\in \mathcal{S}$, and a biased
distribution with $P(A_1A_2)=1-4\protect\alpha k^{-1/2}$, $%
P(A_2A_3)=P(A_3A_4)=P(A_4A_5)=P(A_1A_5)=\protect\alpha k^{-1/2}$, and $%
\protect\alpha=6$. According to Table I, the KCBS violations for the
uniform and biased cases are $\hat{L}^{uni}=3.924$ and $\hat{L}^{bia}=3.905$, respectively. Thus, both $\hat{L}^{uni}$ and $\hat{L}^{bia}$ lie within the interval $[3.9,3.944)$. We plot both min-entropy bounds $kf(\mathcal{L}_m-\epsilon)-\log_2\frac{1}{\delta}$ based respectively on the SDP optimization
and the NS condition (see caption of Fig. 1). The stars represent the
experimental data where approximately $10^5$ measurement outcomes are recorded from the
detectors ($k\approx10^5$). The inset shows the lower bounds of the net entropy
(the output entropy minus the input entropy for distribution $P(A_iA_j)$)
for the biased cases. We get about $5246$ net random numbers from the
experimental data for the biased SDP case. }
\label{NcopyminEntropy}
\end{figure}

\begin{figure*}[tbp]
\begin{center}
\includegraphics[width=16cm]{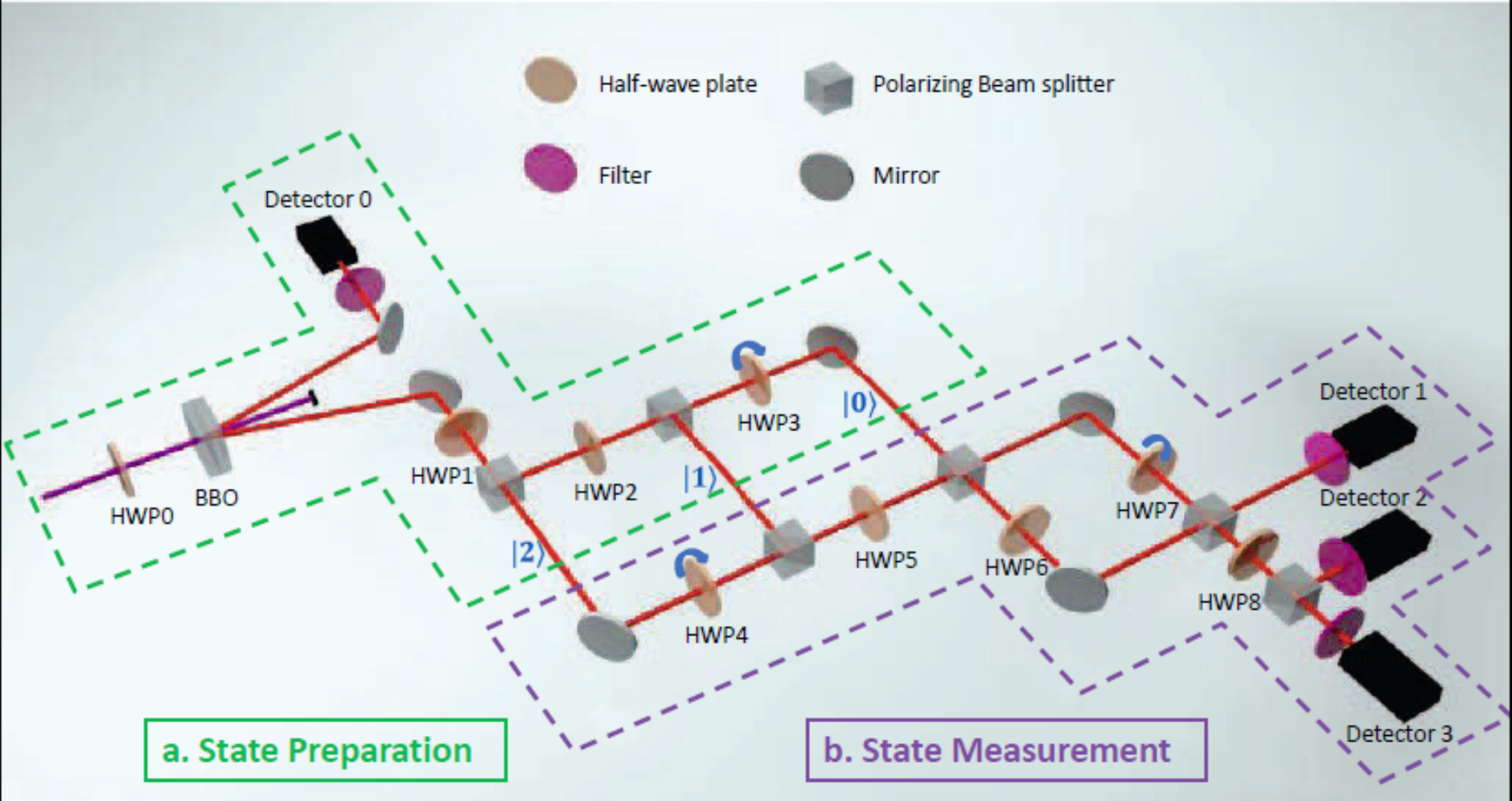}\newline
\caption{Schematic experimental setup of a ture quantum random number
generator. (a) State preparation of a single photonic qutrit. Ultrafast
laser pulses (with a repetition rate of $76$ MHz) at the wavelength of $400$
nm from a frequency doubled Ti:sapphire laser pump two joint
beta-barium-borate (BBO)\ crystals and generate correlated photon pairs at the
wavelength of $800$ nm. A photon-count at the detector D$0$ heralds
a single photon at the other output port, which is split by two polarizing beam splitters (PBS)
into three spatial models, representing a single photonic qutrit. By
adjusting the orientations of the half wave plates (HWP1 and HWP2), we can
prepare any qutrit state. The setup in box (b) implements the
measurements. By tuning the wave plates (HWP5, HWP6, and HWP8), we
measure the probability $P(a_ia_j|A_iA_j)$ for a pair of
compatible observables $(A_i,A_j)$ $(i,j)\in \mathcal{S}$. The input pairs $%
(A_i,A_j)$ are chosen according to the distribution $%
P(A_iA_j) $. To get the desired $(A_i,A_j)$, the orientations of HWPs are listed in the Supporting
information. The wave plates HWP3, HWP4, and HWP7 can be tilted
to balance the Mach-Zender interferometers and their angle are set to
zero. The photons are recorded by single photon detectors D0-D3 after spectra-filters of $3$ nm bandwidth for
coincidence measurements.}
\label{Exp-Setup}
\end{center}
\end{figure*}

Let $\{\mathcal{L}_m: 0\leq m\leq m_{max}\}$ be a series of KCBS violation thresholds with $\mathcal{L}_0=3$ and $\mathcal{L}_{m_{max}}=4\sqrt{5}-5$ corresponding respectively to the classical and quantum bound, and denote $\mathcal{D}(m)$ the probability that the observed KCBS violation $\hat{L}$ lies in the interval $[\mathcal{L}_m,\mathcal{L}_{m+1})$, then we can use the min-entropy to quantify randomness of the output string $\mathcal{O}$~\cite{2010Pironio,2012Pironio,2009Koenig}:
\begin{equation}
E_{\infty }(\mathcal{O}|\mathcal{I},\mathcal{E},m)_{\mathcal{D}}\equiv -\mathtt{log}_{2}\sum_{\mathcal{I},\mathcal{E}}\mathcal{D}(\mathcal{I},\mathcal{E}|m)[\max_{\mathcal{O%
}}\mathcal{D}(\mathcal{O}|\mathcal{I},\mathcal{E},m)],
\end{equation}%
where $\mathcal{E}$ represents the knowledge that a possible adversary has on the state of the device and
 the maximum is taken over all possible values of the output string $%
\mathcal{O}$; the probability distribution $\mathcal{D}(\mathcal{O},\mathcal{I},\mathcal{E})$ is defined in the supplementary information. % and $P(\mathcal{O}|\mathcal{I},\mathcal{E},m)$ is conditional probability
%of obtaining $\mathcal{O}$ conditioned on $\mathcal{I}$ and $\mathcal{E}$ when $\mathcal{L}_m\leq \hat{L}<\mathcal{L}_{m+1}$ has been observed.
In order to
build a link between the KCBS violation and randomness, we assume: (i) the
system can be described by quantum theory; (ii) the input $%
(A_{i_{l}},A_{j_{l}})$ is chosen at step $l$ from an independent random
distribution uncorrelated with the system; (iii) the pair of measurements at
step $l$ are compatible (one measurement does not influence the marginal
distribution of the outcomes of the other measurement); (iv) the adversary's side-information is classical. Based on these
assumptions, we can show that if $\mathcal{D}(m)>\delta$, the min-entropy of the output string conditioned on the input string and the adversary's information has a lower bound (see derivation in Sec. II of the supplementary information):
\begin{equation}
E_{\infty }(\mathcal{O}|\mathcal{I},\mathcal{E},m)_{\mathcal{D}}\geq kf(\mathcal{L}_m-\epsilon)-\log_2\frac{1}{\delta},
\label{min-entropyBound}
\end{equation}%
where the parameter $\epsilon\equiv\{-2[1+(4\sqrt{5}-5)r]^{2}(\ln
\epsilon' )/(kr^{2})\}^{\frac{1}{2}}$ with $r=\min P(A_{i}A_{j})$, the smallest probability of
the input pairs; $\epsilon'$ is another given parameter denoting the closeness between the resulting distribution that characterize $k$ successive use of the device and another extended distribution that is well defined mathematically.  The function $f(L)$ is obtained by semi-definite
programming (SDP)~\cite{1996Vandenberghe} and is shown in Fig. \ref%
{Minentropy}; and the min-entropy bound $kf(\mathcal{L}_m-\epsilon)-\log_2\frac{1}{\delta}$ for
different numbers of trials $k$ is plotted in Fig. \ref{NcopyminEntropy}. It
is remarkable that other than the above four basic assumptions, there is no
further constraint on the states, measurements, or the Hilbert space. It
also requires no assumption that the system behaves identically and
independently for each trial. In particular, the system may have an internal
memory (classical or quantum) so that the results of the $l$th trial depend
on the previous $l-1$ trials. Any observed violation of the KCBS\ inequality
with $\hat{L}>3$ leads to a positive lower bound on the min-entropy, and thus
guarantees genuine randomness generated by the quantum device.

\begin{figure*}[tbp]
\begin{center}
%Requires \usepackage{graphicx}
\includegraphics[width=14cm]{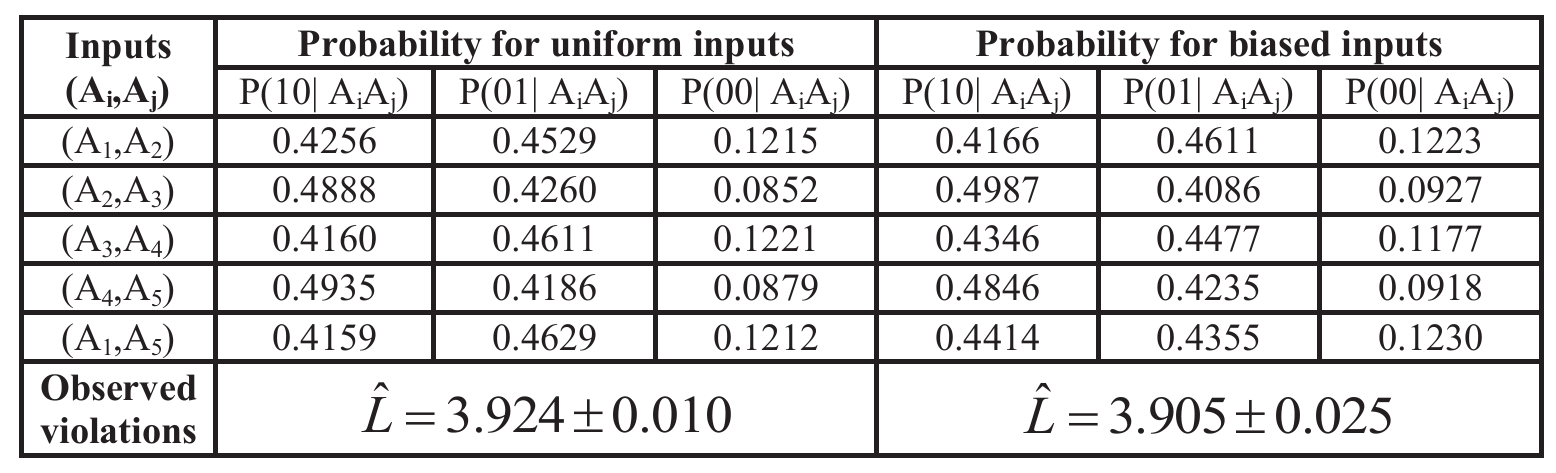}\newline
\caption{Table I: Experimentally observed probabilities and the corresponding KCBS
violation under uniform or biased distribution $P(A_iA_j)$ (see caption of Fig. 2
for specification). The probability $P(11|A_iA_j)$ is negligible as the three-photon coincidence rate
is typically smaller than the two-photon coincidence rate between D0 and Di by more than four orders of
magnitude and thus much less than the error bar. For both cases, the KCBS inequality is significantly
violated, guaranteeing the generation of genuine randomness.}
\label{Table1}
\end{center}
\end{figure*}

In order to experimentally implement our scheme, we use photonic qutrits
where the states are represented by three different paths of a single
photon. For each photonic qutrit, we randomly choose the compatible
measurement configurations $(A_{i}A_{j})$ from the set $\mathcal{S}$
according to a certain probability distribution $P(A_{i}A_{j})$ (uniform or
biased, with its form given in caption of Fig. 2) and record the measurement
outcomes $(a_{i},a_{j})$, which gives our output random bits. To generate random numbers,
we need to observe violation of the KCBS inequality and the
level of violation gives bound on genuine randomness according to Eq. (4).
Different from the experiment in Ref. \cite{2011Lapkiewicz} on test of
quantum contextuality with the KCBS inequality, to generate randomness, the
input pairs $(A_{i},A_{j})$ need to be chosen randomly according to a
probability distribution $P(A_{i}A_{j})$ (instead of fixed before the
experiment), and we need to record the whole measurement output sequence $%
\mathcal{O}=(a_{i_{1}},a_{j_{1}};\cdots ;a_{i_{k}},a_{j_{k}})$ instead of
simply the total number of events $N(a_{i}\neq a_{j}|A_{i}A_{j})$ and $%
N(a_{i}=a_{j}|A_{i}A_{j})$).

The experimental setup is depicted in Fig.~\ref{Exp-Setup}. The spontaneous
parametric down conversion (SPDC) process generates entangled photon pairs.
Through detection of one of the photons by a detector D0, we get a heralded
single-photon source on the other output mode. Two polarization beam
splitters (PBS) split this heralded photon into three spatial models,
representing a single photonic qutrit. Any state of this photonic qutrit can
be prepared by adjusting the orientations of the wave plates before the PBS.
The measurements are implemented by three half wave plates and three
single-photon detectors D1-D3. The angles of these wave plates corresponding
to different pairs of compatible observables are listed in the supporting
information (Table 1 of Sec. IV). We assign value $1$ ($0$) to the
observable $A_{i}$ under a click (non-click) of the corresponding detector.
Due to the inevitable photon loss, there could be no click on the detectors
D1-D3 even when the detector D0 records an event. We discard all the events
in which only the trigger detector D0 and none of the measurement detectors
D1-D3 fires. This is the post-selection technique commonly used in the
photon experiments \cite{2011Lapkiewicz}, which opens up the detection
efficiency loophole. We thus need the fair-sampling assumption that the
photons selected out by the coincidence measurement represent a fair
sampling of all the events. The detection efficiency loophole can be closed
by using single-ion qutrits, where one can follow the same experimental
procedure here and generate true random numbers using only high-speed
single-bit rotations.

The experimental results are summarized in Table I. For both the uniform and
biased input cases, we record about $k\approx 10^{5}$ events. The observed
KCBS violation indicates that $E_{\infty }^{\mathtt{uni}}(\mathcal{O}|\mathcal{I},\mathcal{E},m)>6.3\times 10^{4}$ and $E_{\infty }^{\mathtt{bia}}(\mathcal{O}|%
\mathcal{I})>1.6\times 10^{4}$ with a $99.9\%$ confidence level, so tens of
thousands of genuine random numbers have been generated in both cases.
Similar to Ref.~\cite{2010Pironio}, the scheme described here is actually a
randomness expansion scheme, where a larger random string (the measurement
outcome $\mathcal{O}$) is generated from a smaller set of random seed which
serves as the input $\mathcal{I}$ to specify the measurement configuration $%
(A_{i}A_{j})$. A figure of merit for the randomness expansion scheme is the
net rate of random bits, defined as the number of bits generated minus the
number of bits consumed. In the entanglement based experiment \cite%
{2010Pironio}, it is still difficult to get a positive net rate of random
bits with the current technology because of the slow entanglement generation
rate. In our experiment, for the biased case, we have achieved a positive
net rate for the first time with the output entropy exceeding the input
entropy, leading to approximately $5246$ net random bits. We have performed
extensive random tests on the output strings in our experiment. The results
are summarized in the supplementary information. %The randomness of the
%experimental outcomes is not only supported by passing the known random
%tests, but also guaranteed by certification from the KCBS\ inequality.

We have proposed a scheme to generate genuine random numbers in qutrit
systems where the randomness is guaranteed by violation of the KCBS
inequality, a version of the Kocher-Specker theorem resulting from quantum
contextuality. This scheme guarantees randomness without the need of using
costly quantum resource such as entanglement, and allows for easier
implementation and significantly higher generation rate of random strings.
We have demonstrated this scheme with a proof-of-concept experiment using
photonic qutrits and achieved for the first time a positive net rate of
ture random numbers. The scheme can be readily implemented with other
experimental systems, such as single trapped ions, to close the detection
loophole, opening up practical prospect to generate ture random numbers
with high speeds.

This work was supported by the NBRPC (973 Program)
2011CBA00300 (2011CBA00302) and the NSFC Grant 61033001. DLD and LMD
acknowledge in addition support from the IARPA MUSIQC program, the ARO and
the AFOSR MURI program.

\textit{Note added}.----Having finished this work, we became aware of a recent theoretical work~\cite{2012Abbott}, which explored the Kochen-Specker theorem in another different way to generate randomness.

\section{Supplementary information: Exploring Quantum Contextuality to Generate True Random Numbers}

In this supporting information, we give a detailed derivation of the link
between generation of randomness and violation of the
Klyachko-Can-Binicioglu-Shumovsky (KCBS)\ inequality. For completeness, we
also briefly explain the specific KCBS\ inequality used in our experiment.
On the experimental side, we give detailed configurations of the wave plates
in our experiment and present results for several random tests on the output
data string from the experimental measurements.

\subsection{I. The KCBS inequality}

The KCBS inequality was first introduced in Ref.~\cite{2008KCBS}. It
corresponds to a state-dependent proof of the Kochen-Specker theorem for a
qutrit system. For completeness, here we give a brief derivation. Consider
five two-outcome observables $A_{i}$ $(i=1,2,3,4,5)$ and denote their
outcomes as $a_{i}$, whose values are assigned to be $-1$ or $+1$ (one can
also denote the two outcomes as $0$ and $1$ as in the main text). For any
such assignments, the following algebraic inequality holds: \
\begin{equation}
-a_{1}a_{2}-a_{2}a_{3}-a_{3}a_{4}-a_{4}a_{5}-a_{5}a_{1}\leq 3.  \label{KCBSa}
\end{equation}%
To arrive at the above inequality, we note that the product of the five
monomials on the left-hand side is $-1$. Consequently, at least one term is
equal to $-1$, and the sum of the remaining four terms should not exceed $4$%
. We thus get the above inequality. According to the non-contextual hidden
variable model (NCHVM), if the outcomes $a_{i}$ are described by an unknown
probability distribution, we can integrate over the distribution to take
average and the expectation values of the corresponding observables $A_{i}$
then satisfy
\begin{equation}
-\langle A_{1}A_{2}\rangle -\langle A_{2}A_{3}\rangle -\langle
A_{3}A_{4}\rangle -\langle A_{4}A_{5}\rangle -\langle A_{5}A_{1}\rangle \leq
3,  \label{KCBSb}
\end{equation}
Note that $\langle A_{i}A_{j}\rangle =P(a_{i}=a_{j}|A_{i}A_{j})-P(a_{i}\neq
a_{j}|A_{i}A_{j})$, so the above inequality (\ref{KCBSb}) can also be
written into the following form as shown in the main text:
\begin{equation}
L\equiv \sum_{(i,j)\in \mathcal{S}}[P(a_{i}\neq
a_{j}|A_{i}A_{j})-P(a_{i}=a_{j}|A_{i}A_{j})]\leq 3.  \label{KCBS-Ineq}
\end{equation}%
Any NCHVM should obey the inequality (\ref{KCBS-Ineq}). However, quantum
mechanics violates this inequality for certain measurements on a specific
state. In our experiment, we choose the state to be $\left\vert \Phi
_{0}\right\rangle =|0\rangle $. The five observables are chosen as $%
A_{i}=2|\psi _{i}\rangle \langle \psi _{i}|-\mathbf{I}=2(\alpha
_{i}|0\rangle +\beta _{i}|1\rangle +\gamma _{i}|2\rangle )(\alpha
_{i}\langle 0|+\beta _{i}\langle 1|+\gamma _{i}\langle 2|)-\mathbf{I}$,
where $\mathbf{I}$ is the $3\times 3$ identity matrix and $\alpha
_{1}=\alpha _{2}=\alpha _{3}=\alpha _{4}=\alpha _{5}=\sqrt{\sqrt{5}/5}$, $%
\beta _{1}=-\frac{\sqrt{2}}{2}\cos ^{-1}\frac{\pi }{10}$, $\beta _{3}=\beta
_{4}=-\frac{\sqrt{2}}{2}\tan \frac{\pi }{10}$, $\beta _{2}=\beta _{5}=-\beta
_{1}\cos \frac{\pi }{5}$, $\gamma _{1}=0$, $\gamma _{2}=-\gamma _{5}=\beta
_{1}\sin \frac{\pi }{5}$, and $\gamma _{4}=-\gamma _{3}=-\sqrt{2}/2$. One
can check that all the pairs $(A_{i},A_{j})$ with $(i,j)\in \mathcal{S}$ are
compatible (i.e., $A_{i}$ and $A_{j}$ commute). It is straightforward to
show that for these specific measurements $A_{i}$ under the state $%
\left\vert \Phi _{0}\right\rangle $, quantum mechanics predicts that $L=4%
\sqrt{5}-5\approx 3.944>3$, thus violates the KCBS inequality (\ref%
{KCBS-Ineq}) imposed by any NCHVM.

\subsection{II. Generation of randomness via violation of the KCBS
inequality}

To establish a link between quantum contextuality and randomness, we use
notations and arguments similar to Ref.~\cite{2010Pironio,2012Pironio} for the Bell's
inequality case~\cite{correction}. We say that the observables $O=\{O_{A_{i}}^{a_{i}}\}$ $%
(i=1,2,3,4,5)$ and the state $\rho $ give a quantum realization of the joint
probability $P_{\mathbf{a}_{i}\mathbf{a}_{j}|\mathbf{A}_{i}\mathbf{A}%
_{j}}=\{P(a_{i}a_{j}|A_{i}A_{j})\}$ if $P(a_{i}a_{j}|A_{i}A_{j})=\mathtt{Tr}%
(\rho O_{A_{i}}^{a_{i}}O_{A_{j}}^{a_{j}})$. Here, $O_{A_{i}}^{a_{i}}$ is a
projector that projects the state onto an eigenstate of the observable $A_{i}
$ with eigenvalue $a_{i}$. For simplicity, we denote the quantum realization
and the joint probability distribution as a triplet $\{\rho ,O,P\}$. For one
trial of experiment, the randomness of the output pairs conditioned on the
input pairs $(A_{i},A_{j})$ is defined as the min-entropy:
\begin{equation}
E_{\infty }(\mathbf{a}_{i}\mathbf{a}_{j}|\mathbf{A}_{i}\mathbf{A}_{j})=-\log
_{2}[\max_{a_{i}a_{j}}P(a_{i}a_{j}|A_{i}A_{j})]  \label{Def-Randomness}
\end{equation}%
For any quantum realization of the joint probability $P_{\mathbf{a}_{i}%
\mathbf{a}_{j}|\mathbf{A}_{i}\mathbf{A}_{j}}$ and a given KCBS violation $L$%
, we aim to find a lower bound on the min-entropy:
\begin{equation}
E_{\infty }(\mathbf{a}_{i}\mathbf{a}_{j}|\mathbf{A}_{i}\mathbf{A}_{j})\geq
f(L).  \label{siglef}
\end{equation}%
This is equivalent to solution of the following optimization problem:
\begin{eqnarray}\label{OptProblem}
&\max &\quad \quad \quad P(a_{i}a_{j}|A_{i}A_{j})  \notag \\
&\mathtt{subject}\text{ }\mathtt{to}&\sum_{(i,j)\in \mathcal{S}}[P(a_{i}\neq
a_{j}|A_{i}A_{j})-P(a_{i}=a_{j}|A_{i}A_{j})]=L  \notag \\
&&P(a_{i}a_{j}|A_{i}A_{j})=\mathtt{Tr}(\rho
O_{A_{i}}^{a_{i}}O_{A_{j}}^{a_{j}})
\end{eqnarray}%
where the optimization is carried over all quantum realizations $\{\rho
,O,P\}$. Denote by $P^{\ast }(a_{i}a_{j}|A_{i}A_{j})$ the solution to the
above problem, then the minimal value of $E_{\infty }(\mathbf{a}_{i}\mathbf{a%
}_{j}|\mathbf{A}_{i}\mathbf{A}_{j})$ consistent with the quantum theory and
the KCBS violation $L$ is given by $E_{\infty }(\mathbf{a}_{i}\mathbf{a}_{j}|%
\mathbf{A}_{i}\mathbf{A}_{j})=-\log _{2}P^{\ast }(a_{i}a_{j}|A_{i}A_{j})$.
To get a bound independent of the input pair $(A_{i},A_{j})$, we should
further minimize $E_{\infty }(\mathbf{a}_{i}\mathbf{a}_{j}|\mathbf{A}_{i}%
\mathbf{A}_{j})$ over all the input pairs $(A_{i},A_{j})$. This leads to a
lower bound $f(L)$ on the min-entropy determined by the KCBS violation $L$
only.

The above optimization problem can be efficiently solved by casting it to a
semi-definite programing (SDP) problem. In Refs.~\cite%
{2007Navascues,2008Navascues}, an infinite hierarchy of conditions satisfied
by all quantum correlations are introduced, and the hierarchy is complete in
the asymptotic limit, i.e., if all the conditions in the hierarchy are
satisfied, there always exists a quantum realization $\{\rho ,O,P\}$. All
the conditions in the hierarchy can be transformed to a SDP problem. In
general, conditions higher in the hierarchy are more constraining and thus
give a tighter lower bound $f(L)$ to $E_{\infty }(\mathbf{a}_{i}\mathbf{a}%
_{j}|\mathbf{A}_{i}\mathbf{A}_{j})$. We use the matlab toolboxes SeDuMi~\cite%
{Sturm-Sedumi} to solve the SDP problem for the optimization. The result is
plotted in Fig. 1 of the main text. From the figure, $f(L)$ equals zero at
the classical boundary $L=3$ and increases monotonously as the KCBS
violation $L$ increases. For the maximal violation $L=4\sqrt{5}-5$, $P^{\ast
}=0.457$, corresponding to $f(L)\simeq 1.13$ bits.

We can obtain an upper bound on the min-entropy by numerically searching for
solutions to Eq. (\ref{OptProblem}) under a fixed dimension of the Hilbert
space. When the Hilbert space dimension is fixed to be $3,$ we find that the
upper bound coincides with the lower bound up to a precision of $10^{-5}$.
This indicates that the lower bound obtained above is tight.

The above bound depends on the quantum violation $L$ of the KCBS\
inequality, which itself needs to be determined from a finite runs of
experiments. Now we derive a practical bound on the min-entropy that can be
determined from a finite runs of experiments, taking into account the
statistical error on estimation of the quantum violation $L$ and the classical side information a possible adversary may have on the device.  To this end, let's first introduce the following theorem:

\textbf{Theorem 1}. Suppose we run the experiments  $k$ times and the sequence of inputs $\mathcal{I}=(A_{i_1},A_{j_1};\cdots;A_{i_k},A_{j_k})$ is generated by choosing each pair of inputs $(A_{i_n},A_{j_n})$ independently with probability $P(A_iA_j)$. Let $\delta$, $\epsilon'>0$ be two arbitrary parameters and $r=\min_{ij}\{P(A_{i}A_{j})\}$, then the distribution $P(\mathcal{O}\mathcal{I}\mathcal{E})$ characterizing $k$ successive use of the devices is $\epsilon'$-close to a distribution $\mathcal{D}$ such that, either $\mathcal{D}(m)\leq \delta$ or
\begin{eqnarray}
E_{\infty}(\mathcal{O}|\mathcal{I},\mathcal{E},m)_{\mathcal{D}}\geq kf(\mathcal{L}_m-\epsilon)+\log_2\delta,
\end{eqnarray}
where $\epsilon=(L_q+1/r)\sqrt{-2\ln\epsilon'/k}$ with $L_q=4\sqrt{5}-5$ denoting the maximal KCBS violation.

\textit{Proof}. We follow similar procedures and arguments in Ref.~\cite{2012Pironio} to prove the above theorem.  Define a function $\mathcal{F}(L)=2^{-f(L)}$, then from the solutions to the optimization problem Eq.~\ref{OptProblem} and Fig. 2 in the main text, it is easy to obtain that $\mathcal{F}$ is a concave and monotocially decreasing function. Denote by $\mathcal{O}^{n}=(a_{i_{1}},a_{j_{1}};%
\cdots ;a_{i_{n}},a_{j_{n}})$ ($n\leq k$) the string of outputs before the $(n+1)$th round of experiment (similarly, $%
\mathcal{I}^{n}$ denotes the string of inputs). We define an indicator function $\chi (e)$ as: $\chi
(e)=1$ if the event $e$ happens and $\chi (e)=0$ otherwise. Consider the
following random variable
\begin{eqnarray}\label{randomvar}
\hat{L}_{l}=\sum_{(\mu \nu );(x,y)\in \mathcal{S}}\tau (\mu ,\nu )\frac{\chi
(a_{i_{l}}=\mu ,a_{j_{l}}=\nu ;A_{i_{l}}=x,A_{j_{l}}=y)}{P(xy)},
\end{eqnarray}
where $\mathcal{S}$ is defined in the main text and $\tau (\mu ,\nu )$ $(\mu
,\nu =0,1)$ is a sign function defined as: $\tau (\mu ,\nu )=-1$ if $\mu
=\nu $ and $\tau (\mu ,\nu )=1$ otherwise. It is straightforward to see that
Eq. (\ref{randomvar}) corresponds to the KCBS expression (2) in the main
text and the expectation value of $\hat{L}_{l}$ conditional on $W^{l}$ is equal to $L(W^{l})$, i.e., $\mathbb{E}(\hat{L}|W^l)=L(W^{l})$. Here $\mathcal{W}^{l}\equiv(\mathcal{O}^{l-1}\mathcal{I}^{l-1}\mathcal{E})$ denotes all the
events before the $l$th round of experiment and the possible adversary's classical side information. Let $\hat{L}=\frac{1}{k}\sum_{l=1}^{k}\hat{L}_{l}$ be our estimator of the
KCBS violation. After specify the above notations, now let's also introduce two lemmas for the proof of the theorem:

\textbf{Lemma 1}. For a given parameter $\epsilon'>0$, let $\epsilon=(L_q+1/r)\sqrt{-2\ln\epsilon'/k}$ and $\mathcal{T}_{\epsilon}=\{(\mathcal{O},\mathcal{I},\mathcal{E})|\frac{1}
{k}\sum_{l=1}^k\mathbb{E}(\hat{L}_l|W^l)\geq \hat{L}(\mathcal{O},\mathcal{I})-\epsilon\}$, then we have:

(i) for any $(\mathcal{O},\mathcal{I},\mathcal{E})\in \mathcal{T}_{\epsilon}$,
\begin{eqnarray}\label{POIE}
P(\mathcal{O}|\mathcal{I}\mathcal{E})\leq \mathcal{F}^k(\hat{L}(\mathcal{O},\mathcal{I})-\epsilon).
\end{eqnarray}

(ii)
\begin{eqnarray}\label{PrT1}
\texttt{Pr}(\mathcal{T}_{\epsilon})=\sum_{(\mathcal{O},\mathcal{I},\mathcal{E})\in \mathcal{T}_{\epsilon}}P(\mathcal{O},\mathcal{I},\mathcal{E})\geq 1-\epsilon'.
\end{eqnarray}

\textit{Proof}. By using the Bayes's rule and the fact that the response of the system does not depend on the future inputs and outputs, we have:

\begin{eqnarray}
P(\mathcal{O}|\mathcal{I}\mathcal{E}) &=&\prod_{l=1}^{k}P(a_{i_{l}}a_{j_{l}}|\mathcal{O}^{l-1}\mathcal{I}^{l}\mathcal{E})
\notag  \label{IntroWi} \\
&=&\prod_{l=1}^{k}P(a_{i_{l}}a_{j_{l}}|A_{i_{l}}A_{j_{l}}\mathcal{W%
}^{l})
\end{eqnarray}
From Eq.~\ref{OptProblem}, the probability $%
P(a_{i_{l}}a_{j_{l}}|A_{i_{l}}A_{j_{l}}\mathcal{W}^{l})$ is bounded by a
function of the KCBS violation $L(W^{l})$: $%
P(a_{i_{l}}a_{j_{l}}|A_{i_{l}}A_{j_{l}}\mathcal{W}^{l})\leq \mathcal{F}(L(W^{l}))$. Thus, we have:
\begin{eqnarray}
P(\mathcal{O}|\mathcal{I}\mathcal{E}) &\leq& \prod_{l=1}^{k}\mathcal{F}(L(W^{l}))\nonumber\\
&\leq&\mathcal{F}^k (\frac{1}{k}\mathbb{E}(\hat{L}_l|W^l))\nonumber\\
&\leq&\mathcal{F}^k(\hat{L}(\mathcal{O},\mathcal{I})-\epsilon),
\end{eqnarray}
where the equality $\mathbb{E}(\hat{L}|W^l)=L(W^{l})$ and the fact that $\mathcal{F}$ is logarithmically concave are used in the second inequality. For the third inequality, we used the definition of $\mathcal{T}_{\epsilon}$ and the fact that $\mathcal{F}$ is monotonically decreasing.

To prove Eq.(\ref{PrT1}), let's define another random variable $M^{q}=\sum_{l=1}^{q}(%
\hat{L}_{l}-\mathbb{E}(\hat{L}|W^l))$. The sequence $\{M^{q}:q\geq 1\}$ is a martingale
process~\cite{2001Grimmett-Book}. The range of the martingale increment is
bounded by $|\hat{L}_{l}-L(W^{l})|\leq \frac{1}{r}+L_q$. From the Azuma-Hoeffding inequality $P(M^{q}\geq k\epsilon)\leq \exp (-\frac{(k\epsilon)^2}{2k(1/r+L_q)^{2}})$~%
\cite{1967Azuma,1960Hoeffding,2001Grimmett-Book}, we have
\begin{equation}
P\left( \frac{1}{k}\sum_{l=1}^{k}\mathbb{E}(\hat{L}|W^l)\leq \frac{1}{k}\sum_{l=1}^{k}\hat{L%
}_{l}-\epsilon \right) \leq \epsilon',  \label{MargEq}
\end{equation}%
where the equation
$\epsilon=(L_q+1/r)\sqrt{-2\ln\epsilon'/k}$ is used. Eq. (\ref{MargEq}) combined with the definition of $\mathcal{T}_{\epsilon}$ gives the Eq.~(\ref{PrT1}) desired.

The above discussion considered the case that the random variable sequence $\mathcal{O}$ only takes values in the output space $\mathbb{S}^k=\{-1,1\}^k$. Similar as in Ref.~\cite{2012Pironio}, we extend the range of $\mathcal{O}$ and view it as an element of $\mathbb{S}^k\cup\bot$ with $P(\mathcal{O}|\mathcal{I}\mathcal{E})=0$ if $\mathcal{O}=\bot$. In fact, $\bot$ can be regarded as an ``abort-output" produced by the devices, from which no KCBS violation has be obtained.

\textbf{Lemma 2}. There exists a probability distribution $\mathcal{D}=\{\mathcal{D}(\mathcal{O},\mathcal{I},\mathcal{E})\}$ that is $\epsilon'$-close to $P=\{P(\mathcal{O},\mathcal{I},\mathcal{E})\}$, i.e., $d(\mathcal{D},P)=\frac{1}{2}\sum_{\mathcal{O},\mathcal{I},\mathcal{E}}|
P(\mathcal{O},\mathcal{I},\mathcal{E})-\mathcal{D}(\mathcal{O},\mathcal{I},\mathcal{E})|\leq \epsilon'$. Distribution $\mathcal{D}$ also satisfy the condition:
\begin{eqnarray}\label{Dcondition2}
\mathcal{D}(\mathcal{O}|\mathcal{I},\mathcal{E})\leq \mathcal{F}^k(\hat{L}(\mathcal{O},\mathcal{I})-\epsilon),
\end{eqnarray}
for all $(\mathcal{O},\mathcal{I},\mathcal{E})$ such that $\mathcal{O}\neq\bot$.

\textit{Proof}. We only have to construct a probability distribution satisfy all the conditions. Let $\mathcal{D}(\mathcal{O},\mathcal{I},\mathcal{E})=P(\mathcal{I})P(\mathcal{E})\mathcal{D}(\mathcal{O}|\mathcal{I},\mathcal{E})$, with $\mathcal{D}(\mathcal{O}|\mathcal{I},\mathcal{E})$ defined as:
(i) $\mathcal{D}(\mathcal{O}|\mathcal{I},\mathcal{E})=P(\mathcal{O}|\mathcal{I},\mathcal{E})$ if $(\mathcal{O},\mathcal{I},\mathcal{E})\in \mathcal{T}_{\epsilon}$; (ii) $\mathcal{D}(\mathcal{O}|\mathcal{I},\mathcal{E})=0$ if $\mathcal{O}\neq\bot$ and $(\mathcal{O},\mathcal{I},\mathcal{E})\notin\mathcal{T}_{\epsilon}$; (iii) $\mathcal{D}(\bot|\mathcal{I},\mathcal{E})=1-
\sum_{(\mathcal{O},\mathcal{I},\mathcal{E})\notin\mathcal{T}_{\epsilon}}P(\mathcal{O}|\mathcal{I},\mathcal{E})$.
Then it is straightforward to obtain from Lemma 1 that $\mathcal{D}$ satisfies Eq.~(\ref{Dcondition2}) for all $(\mathcal{O},\mathcal{I},\mathcal{E})$ such that $\mathcal{O}\neq\bot$, and
\begin{eqnarray}
&&d(\mathcal{D},P)\nonumber\\
&=&\frac{1}{2}\sum_{\mathcal{O},\mathcal{I},\mathcal{E}}|
P(\mathcal{O},\mathcal{I},\mathcal{E})-\mathcal{D}(\mathcal{O},\mathcal{I},\mathcal{E})|\nonumber\\
&=&\frac{1}{2}\sum_{\mathcal{I},\mathcal{E}}P(\mathcal{I},\mathcal{E})\sum_{\mathcal{O}}
|P(\mathcal{O}|\mathcal{I},\mathcal{E})-\mathcal{D}(\mathcal{O}|\mathcal{I},\mathcal{E})|\nonumber\\
&=&\frac{1}{2}[\sum_{(\mathcal{O},\mathcal{I},\mathcal{E})\notin\mathcal{T}_{\epsilon}}P(\mathcal{O},\mathcal{I},\mathcal{E})
+1-\sum_{(\mathcal{O},\mathcal{I},\mathcal{E})\in\mathcal{T}_{\epsilon}}P(\mathcal{O},\mathcal{I},\mathcal{E})]\nonumber\\
&\leq&\epsilon'.\nonumber
\end{eqnarray}

After introducing the above two lemmas, now we are ready to prove Theorem 1. As in the main text, let $\{\mathcal{L}_m: 0\leq m\leq m_{max}\}$ be a series of KCBS violation shresholds and $\mathcal{D}(m)$ the probability that the observed KCBS violation $\hat{L}$ lies in the interval $[\mathcal{L}_m,\mathcal{L}_{m+1})$. Denote $Y_m=\{\mathcal{O}|\mathcal{O}\neq\bot\; \texttt{and}\;\mathcal{L}_m\leq\hat{L}<\mathcal{L}_{m+1}\}$. By using Lemma 2 and the fact that $\mathcal{F}$ is monotically decreasing, we have:
\begin{eqnarray}
&&E_{\infty }(\mathcal{O}|\mathcal{I},\mathcal{E},m)_{\mathcal{D}}\nonumber\\
&\equiv& -\mathtt{log}_{2}\sum_{\mathcal{I},\mathcal{E}}\mathcal{D}(\mathcal{I},\mathcal{E}|m)[\max_{\mathcal{O%
}}\mathcal{D}(\mathcal{O}|\mathcal{I},\mathcal{E},m)]\nonumber\\
&=&-\mathtt{log}_{2}\sum_{\mathcal{I},\mathcal{E}}\mathcal{D}(\mathcal{I},\mathcal{E}|m)\frac{1}{\mathcal{D}(m|\mathcal{I},\mathcal{E})}
\max_{\mathcal{O}\in Y_m}\mathcal{D}(\mathcal{O}|\mathcal{I},\mathcal{E})\nonumber\\
&\geq& -\mathtt{log}_{2}\sum_{\mathcal{I},\mathcal{E}}\mathcal{D}(\mathcal{I},\mathcal{E}|m)\frac{\mathcal{F}^k(\mathcal{L}_m-\epsilon)}{\mathcal{D}(m|\mathcal{I},\mathcal{E})}\nonumber\\
&=&-\mathtt{log}_{2}\sum_{\mathcal{I},\mathcal{E}}\frac{\mathcal{D}(\mathcal{I},\mathcal{E})}{\mathcal{D}(m)}\mathcal{F}^k(\mathcal{L}_m-\epsilon)\nonumber\\
&=& kf(\mathcal{L}_m-\epsilon)-\log_2\frac{1}{\mathcal{D}(m)}.\nonumber
\end{eqnarray}
Here in the last inequality, the equation $f=-\log_2\mathcal{F}$ is used. The above equation immediately leads to the claims in Theorem 1.

Theorem 1 tells us that there is essentially no difference between the distribution $P$, which characterize the outputs $\mathcal{O}$ of the devices and their correlations with the inputs $\mathcal{I}$ and the adversary's classical side information $\mathcal{E}$, and the distribution $\mathcal{D}$ defined above~\cite{2012Pironio}. If we have confidence that the observed KCBS violation $\hat{L}$ lies in $[\mathcal{L}_m,\mathcal{L}_{m+1})$ with non-negligible probability, i.e., $\mathcal{D}(m)>\delta$, then the entropy of the outputs $\mathcal{O}$ is guaranteed to have a positive lower bound $kf(\mathcal{L}_m-\epsilon)-\log_2\frac{1}{\delta}$, that is, the randomness of the outputs is guaranteed to be larger than $kf(\mathcal{L}_m)$ up to epsilonic corrections.

\subsection{III. Generation of randomness under relaxed conditions}

It has been shown in Ref.~\cite{2010Pironio} that violation of Bell's
inequality can be used to certify randomness even without the need of
quantum mechanics. One only needs to assume the no-signalling (NS)
condition: for two measurements corresponding to space-like events, one
measurement has no influence on the marginal distribution of the outcomes of
the other measurement. Here, for the single qutrit protocol, we can
similarly assume a relaxed condition that corresponds to the NS condition
for bipartite systems. For two compatible measurements, we can assume one
measurement has no influence on the marginal distribution of the outcomes of
the other measurement. Quantum mechanics obviously obey this rule. So,
compared with the assumption of full formalism of quantum mechanics, this
condition corresponds to a significantly relaxed requirement. To emphasize
the correspondence, we still call this assumption the NS condition, although
it is not directly connected with no signaling for single qutrit systems.
Under only the NS condition, the optimization problem (\ref{OptProblem})
should be replaced by
\begin{eqnarray}
&\max &\quad \quad \quad P(a_{i}a_{j}|A_{i}A_{j})  \notag \\
&\mathtt{subject}\text{ }\mathtt{to}&\sum_{(i,j)\in \mathcal{S}}[P(a_{i}\neq
a_{j}|A_{i}A_{j})-P(a_{i}=a_{j}|A_{i}A_{j})]=L  \notag \\
&&0\leq P(a_{i}a_{j}|A_{i}A_{j})\leq 1, \\
&&\sum_{a_{i}a_{j}}P(a_{i}a_{j}|A_{i}A_{j})=1,  \notag \\
&&\sum_{a_{j}}P(a_{i}a_{j}|A_{i}A_{j})=P(a_{i}|A_{i}),  \notag \\
&&\sum_{a_{i}}P(a_{i}a_{j}|A_{i}A_{j})=P(a_{j}|A_{j}),  \notag
\end{eqnarray}%
where the last two equalities are mathematical description of the NS
condition. With a given quantum violation $L$ of the KCBS inequality, we can
analytically solve the above optimization problem using linear programming
and obtain $f(L)=-\log _{2}(1.75-L/4)$. In Fig. 1 of the main text, we plot
this analytic bound $f(L)$ versus $L$ under the NS condition. Its value
becomes strictly positive as soon as $L$ exceeds the classical bound $3$.

\subsection{IV. Experimental configuration of the wave plates}

In this section, we give more details on the experimental configuration of
the half wave plates. The experiment setup is show in Fig. 3 of the main
text. As stated in section I, we choose the qutrit state to be $\Phi
_{0}=|0\rangle $, which is prepared by setting the angles of HWP0, HWP1 and
HWP2 to be $0$, $\pi /4$, and $-\pi /4$, respectively. Using the linear
optics transformation rules for the HWPs and the PBS, we find that for this
setup a click in the detector D1 (D2) corresponds respectively to a
projection to the state $|\psi _{1}\rangle $ ($|\psi _{2}\rangle $), with $%
|\psi _{1}\rangle =\cos (2\theta _{2})|0\rangle -\sin (2\theta _{2})\cos
(2\theta _{1})|1\rangle -\sin (2\theta _{2})\sin (2\theta _{1})|2\rangle $
and $|\psi _{2}\rangle =\cos (2\theta _{3})\sin (2\theta _{2})|0\rangle
+[\cos (2\theta _{3})\cos (2\theta _{2})\cos (2\theta _{1})-\sin (2\theta
_{3})\sin (2\theta _{1})]|1\rangle +[\cos (2\theta _{3})\cos (2\theta
_{2})\sin (2\theta _{1})+\sin (2\theta _{3})\cos (2\theta _{1})]|2\rangle $.
Here, $\theta _{1}$, $\theta _{2}$ and $\theta _{3}$ denote the angles of
HWP5, HWP6, and HWP8, respectively. Based on this transformation, we obtain
the angles of the HWPs corresponding to the measurements $A_{i}$ given in
Sec. I of this supplementary information. These angles and the their
corresponding observables are listed in Table.~\ref{HWP-Angle}.

\begin{table}[tbp]
%Requires \usepackage{graphicx}
\includegraphics[width=80mm]{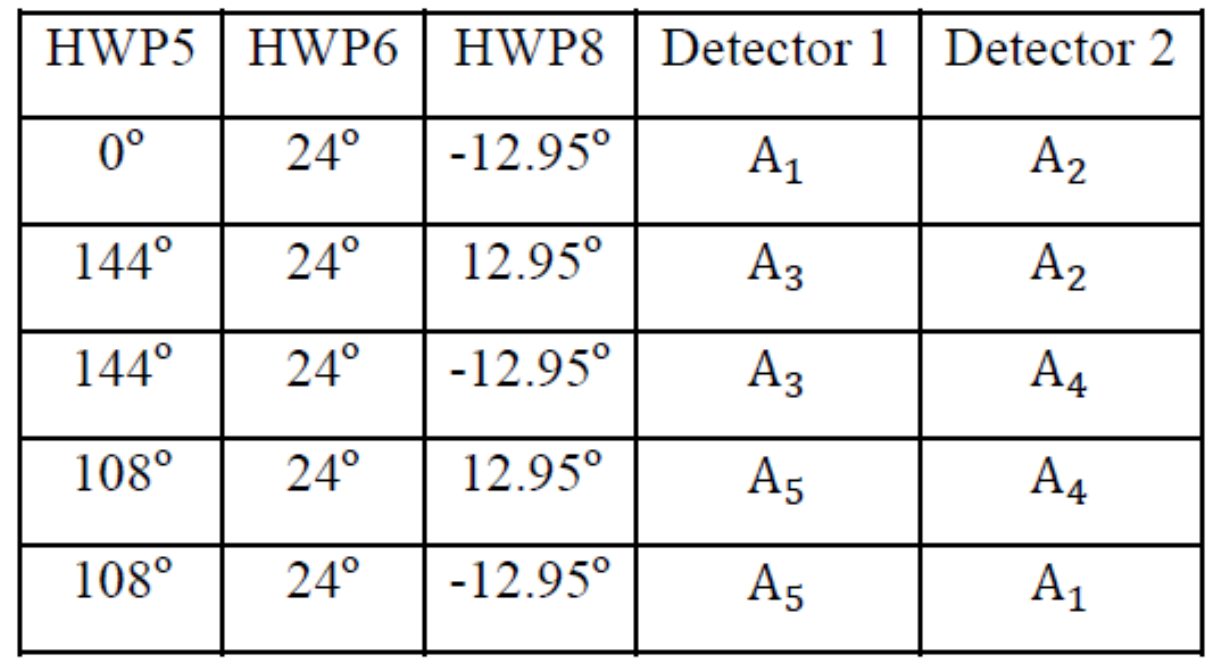}\newline
\caption{$|$ Angles of the half-wave plates (HWP5, HWP6 and HWP8 in Fig. 3
of the main text) to measure five pairs of compatible observables. }
\label{HWP-Angle}
\end{table}

\subsection{V. Statistical tests of the generated random numbers}

To check the quality of the random numbers generated in our experiment, we
carry out a number of statistical random tests~\cite%
{NIST-RandomTest,1996Menezes}. The length of the output string in our
experiment is about $10^{5}$, so we choose the random tests that are
statistically relevant at this string size. To be specific, we perform the
random tests called "Frequency", "Block Frequency", "Runs",
"Longest-Run-of-Ones in a Block (LROB)", "Non-overlapping Template Matching
(NOTM)", "Serial", "Approximate Entropy (AE)", "Cumulative Sums (Cusums)"~%
\cite{NIST-RandomTest}, and "Two-bit"~\cite{1996Menezes}. All these tests
are implemented by Mathematica programs. For the qutrit system, quantum
theory predicts that the number of ones in our output strings should be
larger than that of zeros in the output string. So, we first perform a Von
Neumann extractor~\cite{1951Neumann} to the rough data before the tests.

The test results are summarized in Table II. What we show in the table is
the so-called p-values, which are indicators of the test results. More
precisely, a p-value is the probability that an idea random number generator
would have produced a sequence less random than the sequence in test~\cite%
{2010Pironio,NIST-RandomTest}. In other words, a bigger p-value indicates
that the sequence in test is more likely to be random. Therefore, a p-value
of $0$ simply means that the tested sequence appears to be completely
non-random, whereas a p-value of $1$ implies that the sequence in test
appears to be perfectly random. Usual p-values lies in the open interval $%
(0,1)$ and a significance level $\vartheta $ should be introduced for the
test. If the p-value $\geq \vartheta $, we accept the tested sequence as
random. Otherwise, it is non-random. Typically, $\vartheta $ is chosen to be
in the range $[0.0001,0.01]$. Here, we choose $\vartheta =0.001$. A sequence
with a p-value larger than $0.001$ passes the test and is considered to be a
random sequence, otherwise it fails the test.

From the Table, all the four sequences generated by the detector D1 and D2
separately pass all the tests. This confirms the validity of the experiment.
For both the uniform and the biased input cases, the joint output strings
produced by the detectors D1 and D2 arranged in the order $%
(a_{i_{1}},a_{j_{1}};\cdots ;a_{i_{k}},a_{j_{k}})$ cannot pass the test.
This is expected since the measurement outputs of the D1 and D2 detectors
are correlated due to quantum contextuality. We should note that the random
tests just confirm our expectation. No random tests on finite strings should
be considered complete. Much stronger evidence of randomness in the output
string of our experiment is provided by the observed
KCBS violation, which is independent of any hypothesis on how the experiment
was carried out. Violation of the KCBS inequality guarantees that the
entropy of output string has a positive lower bound. In this case, one can
always use a randomness extractor~\cite{1951Neumann,1999Nisan} to convert
the string into a new one of size $kf(\mathcal{L}_m-\varepsilon )$, which is
almost uniformly distributed and perfectly random.

\begin{table}[tbp]
%Requires \usepackage{graphicx}
\includegraphics[width=85mm]{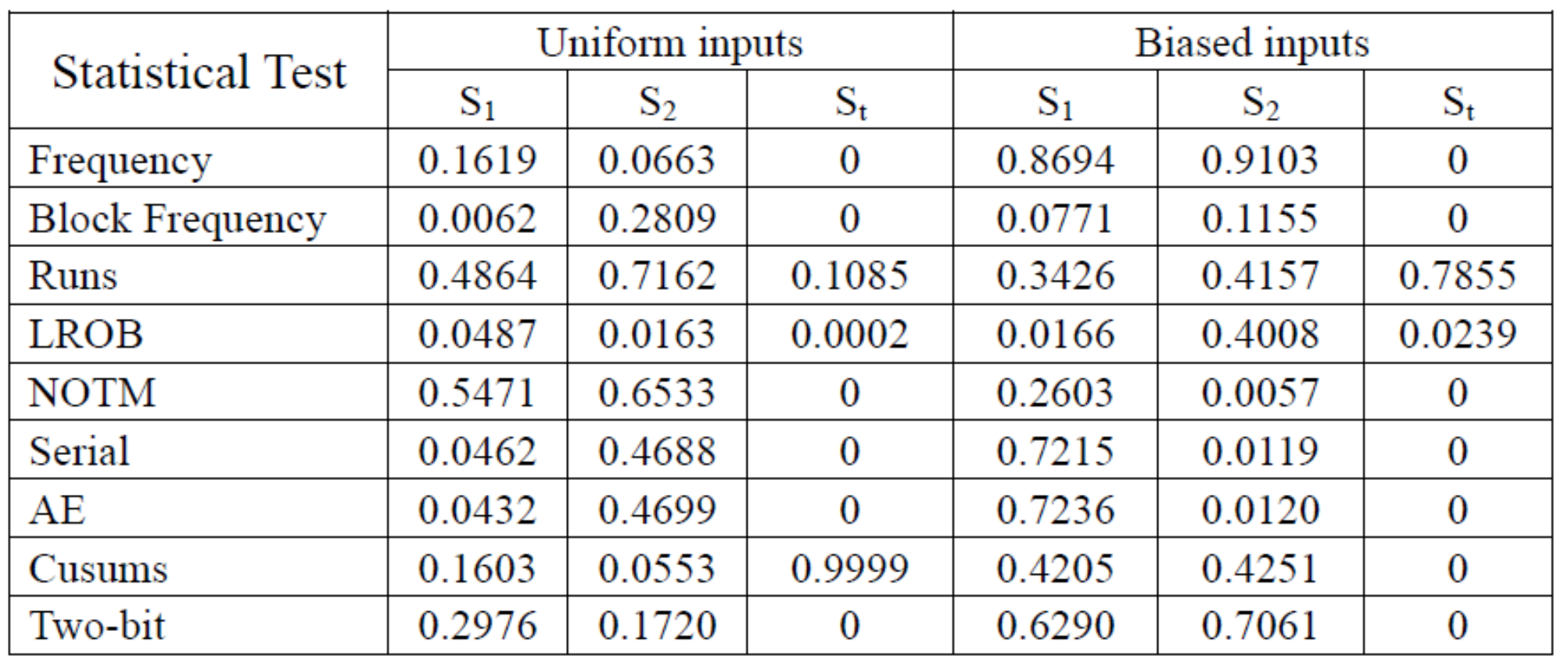}\newline
\caption{$|$ Results of the random tests described by the p-values of the
output strings. $S_1$ and $S_2$ denote the sequences separately generated by
detector D1 and D2, respectively. $S_t$ is the output string of D1 and D1
together, arranged in the order $(a_{i_0},a_{j_0};\cdots;a_{i_k},a_{j_k})$,
which has finite correlation due to quantum contextuality and cannot pass
any of the random tests that are sensitive to correlation.}
\label{RandomTest-Restults}
\end{table}

\end{document}